\documentclass[11pt,a4paper,pdftex]{article}
\pdfoutput=1
\usepackage{jcappub}

\newcommand{\be}{\begin{eqnarray}}
\newcommand{\ee}{\end{eqnarray}}
\newcommand{\rar}{\rightarrow}

\begin{document}

\title{Constraining possible variations\\ 
of the fine structure constant\\ 
in strong gravitational fields\\
with the K$\alpha$ iron line}

\author{Cosimo Bambi}

\emailAdd{bambi@fudan.edu.cn}

\affiliation{Center for Field Theory and Particle Physics \& Department of Physics,\\
Fudan University, 220 Handan Road, 200433 Shanghai, China}

\abstract{In extensions of general relativity and in theories aiming at unifying gravity 
with the forces of the Standard Model, the value of the ``fundamental constants'' 
is often determined by the vacuum expectation value of new fields, which may thus 
change in different backgrounds. Variations of fundamental constants with respect 
to the values measured today in laboratories on Earth are expected to be more 
evident on cosmological timescales and/or in strong gravitational fields. In this 
paper, I show that the analysis of the K$\alpha$ iron line observed in the X-ray 
spectrum of black holes can potentially be used to probe the fine structure constant 
$\alpha$ in gravitational potentials relative to Earth of $\Delta \phi \approx 0.1$. 
At present, systematic effects not fully under control prevent to get robust and
stringent bounds on possible variations of the value of $\alpha$ with
this technique, but the fact that current data can be fitted with models based 
on standard physics already rules out variations of the fine structure constant 
larger than some percent.}


\maketitle


\section{Introduction}

The Einstein Equivalence Principle (EEP) is a fundamental concept in general 
relativity and in other theories of gravity. It is based on the following three 
assumptions~\cite{will}:
\begin{enumerate}
\item Weak Equivalence Principle (WEP). The trajectory of a freely-falling 
test-particle is independent of its internal structure and composition.
\item Local Lorentz Invariance (LLI). The result of any local non-gravitational 
experiment is independent of the velocity of the freely-falling reference frame 
in which it is performed.
\item Local Position Invariance (LPI). The result of any local non-gravitational 
experiment is independent of where and when in the Universe it is performed.
\end{enumerate}
The WEP is equivalent to the statement that the ``inertial mass'' is always 
proportional to the ``gravitational mass'', and therefore we can choose a system
of units in which such a proportionality constant is 1. Moreover, it seems (Schiff's 
conjecture) that any complete and self-consistent theory of gravity in which the
WEP holds necessarily implies the LLI and the LPI, and therefore the EEP~\cite{will}.
However, a rigorous proof on such a conclusion is lacking. The EEP is 
satisfied by metric theories of gravity, including general relativity and all those
frameworks in which the spacetime is characterized by a symmetric metric 
and possible new degrees of freedom universally couple to matter.

Extensions of general relativity and theories aiming at unifying gravity with the 
forces of the Standard Model typically violate the 
EEP~\cite{jor,bra,mar,ysw,mae,tay,dam1,dam2,dam3} (see also Appendix~\ref{app-a}). 
Tests of the EEP can thus 
be seen as an alternative strategy to discover new interactions and they can 
potentially find signatures of high energy physics in the low energy gravity sector.
The search for time and/or space variations of fundamental constants is a quite 
hot topic today and can involve different research fields~\cite{uzan1,uzan2}. It is 
a test of the LPI, and therefore of the EEP. For instance, in superstring theories, 
the low energy parameters of the Standard Model depend on the vacuum 
expectation value of scalar fields (dilaton, moduli). As the latter is determined 
by the background geometry, it is quite natural to expect both temporal variation 
on cosmological scales and a dependence on the gravitational potential of the 
values of these quantities. However, clear predictions are not possible, and 
the common strategy is to adopt a phenomenological approach, constraining 
possible variations of fundamental constants in many different ways.

The possibility of time and/or space variations of the fine structure constant
$\alpha = e^2 / \hbar c \approx 1/137$ has attracted a lot of interest.
Atomic clock experiments can constrain possible temporal variations of $\alpha$ 
today at the level of $|\dot{\alpha}/\alpha| \lesssim 10^{-17}$~yr$^{-1}$~\cite{clock}.
In Ref.~\cite{webb1a,webb1b}, the authors reported evidence for a different value 
of $\alpha$ in quasar spectra at high redshift, which was initially interpreted as
an indication for a temporal variation of the fine structure constant.
More recent studies seem instead to point out a spacial variation of $\alpha$,
with the existence of a preferred direction in the Universe: for redshift $z > 1.8$,
data from the Keck telescope for the Northern sky suggest $\Delta\alpha/\alpha 
= (-0.74\pm0.17) \cdot 10^{-5}$, while data from the VLT telescope for the 
Southern sky give $\Delta\alpha/\alpha = (+0.61\pm0.20) \cdot 10^{-5}$~\cite{webb2}. 
While it is not clear how robust these results are, they cannot be explained 
by the known systematic effects. The dependence of the fine structure constant 
on strong gravitational fields has been studied in~\cite{berengut}. The authors 
compare laboratory spectra with far-UV astronomical spectra from the white-dwarf 
star G191-B2B and obtain the constraints
\be
\Delta\alpha/\alpha &=& (4.2\pm1.6) \cdot 10^{-5} 
\quad ({\rm FeV}) \, , \nonumber\\
\Delta\alpha/\alpha &=& (-6.1\pm5.8) \cdot 10^{-5} 
\quad ({\rm NiV}) \, ,
\ee
for a dimensionless gravitational potential relative to Earth of $\Delta\phi \approx 
5 \cdot 10^{-5}$, where $\phi = G_{\rm N} M / r c^2$, $M$ is the mass of the white 
dwarf/Earth, and $r$ its radius.

\section{K$\alpha$ iron line}

The X-ray spectrum of black hole candidates is often characterized by the 
presence of a power-law component. This feature is commonly interpreted as 
the inverse Compton scattering of thermal photons by electrons in a hot corona 
above the accretion disk. This ``primary component'' irradiates also the accretion 
disk, producing a ``reflection component'' and spectral lines by fluorescence
in the X-ray spectrum. The strongest line is the K$\alpha$ iron line at 6.4~keV. 
This line is intrinsically narrow in frequency, while the one observed appears 
broadened and skewed. The interpretation is that the line is strongly altered 
by special and general relativistic effects, which produce a characteristic profile, 
first predicted and identified from Cygnus~X-1 data in Ref.~\cite{ka1} (though that 
observation was originally published in~\cite{ka2}). For a review, see e.g. 
Refs.~\cite{ka-rev1,ka-rev2}.

The profile of the K$\alpha$ iron line depends on the background metric, the 
geometry of the emitting region, the disk emissivity, and the diskÕs inclination 
angle with respect to the line of sight of the distant observer. In 4-dimensional
general relativity, black holes are described by the Kerr solution and are
characterized by only two parameters; that is, the mass $M$ and the spin angular 
momentum $J$. The dimensionless spin parameter $a_* = J/M^2$ is the only
relevant parameter of the background geometry, while $M$ sets the length of 
the system, without affecting the line profile. In those sources for which there 
is indication that the line is mainly emitted close to the compact object, the 
emission region is thought to range from the radius of the innermost stable 
circular orbit (ISCO), $r_{\rm in} = r_{\rm ISCO}$, to some outer radius $r_{\rm out}$. 
The disk emissivity is often assumed to be a power-law in radius of the form
$I_{\rm e} \propto 1/r^q$. The simple lamp-post model gives $q = 3$ at large
radii $r \gg r_g$, where $r_g = G_{\rm N}M/c^2$ is the gravitational radius of the 
object. This becomes a quite bad approximation for the inner accretion disk,
which can be very important for the spin measurement of fast-rotating black holes. 
In Ref.~\cite{dauser}, the authors compute the emissivity index $q$ in the lamp-post 
model as a function of the radial coordinate for different lamp-post heights (see their 
Fig.~3). In practice, one often assumes an inner and an outer emissivity indices, 
say $q_1$ and $q_2$, which are determined during the fitting procedure~\cite{fab12}. 
The fourth parameter is the inclination of the disk with respect to the line of sight 
of the distant observer, $i$. The dependence of the line profile on $a_*$, $i$, 
$q$, and $r_{\rm out}$ in the Kerr background has been analyzed in detail 
by many authors, starting from Ref.~\cite{ka1}. The profile of the K$\alpha$ 
iron line in non-Kerr spacetimes is discussed in 
Refs.~\cite{torres,iron1,iron2,iron3,iron4}.

\begin{figure}
\begin{center}
\includegraphics[type=pdf,ext=.pdf,read=.pdf,width=7.5cm]{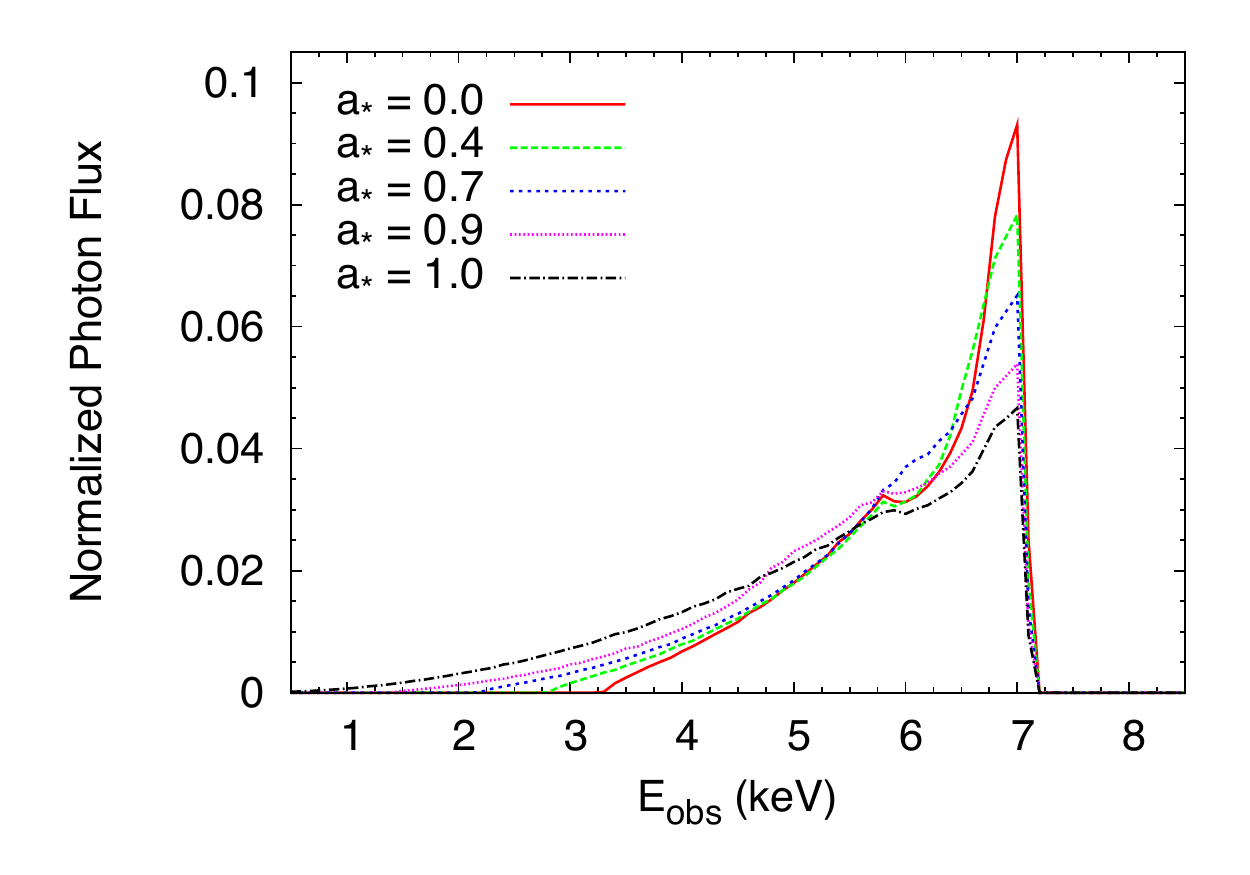}
\includegraphics[type=pdf,ext=.pdf,read=.pdf,width=7.5cm]{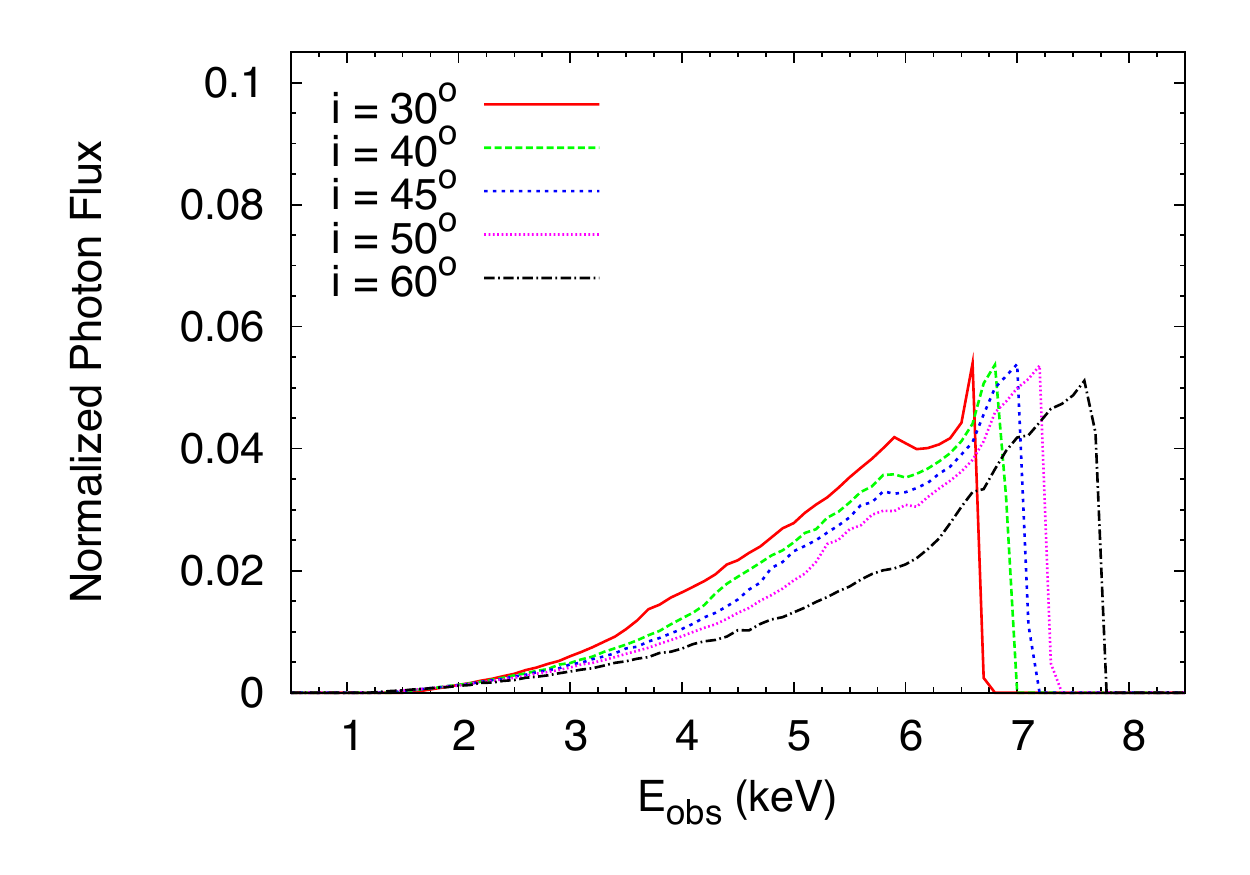}
\end{center}
\vspace{-0.5cm}
\caption{Left panel: K$\alpha$ iron line in Kerr spacetime with different values of 
the spin parameter $a_*$ and viewing angle $i = 45^\circ$. Right panel: K$\alpha$ 
iron line in Kerr spacetime with spin parameter $a_* = 0.9$ and different values
of the viewing angle $i$. The other parameters of the model are: inner radius 
$r_{\rm in} = r_{\rm ISCO}$, outer radius $r_{\rm out} = r_{\rm in} + 100 \, r_g$, index
of the emissivity profile $q = 3$. \label{f1}}
\vspace{1.0cm}
\begin{center}
\includegraphics[type=pdf,ext=.pdf,read=.pdf,width=7.5cm]{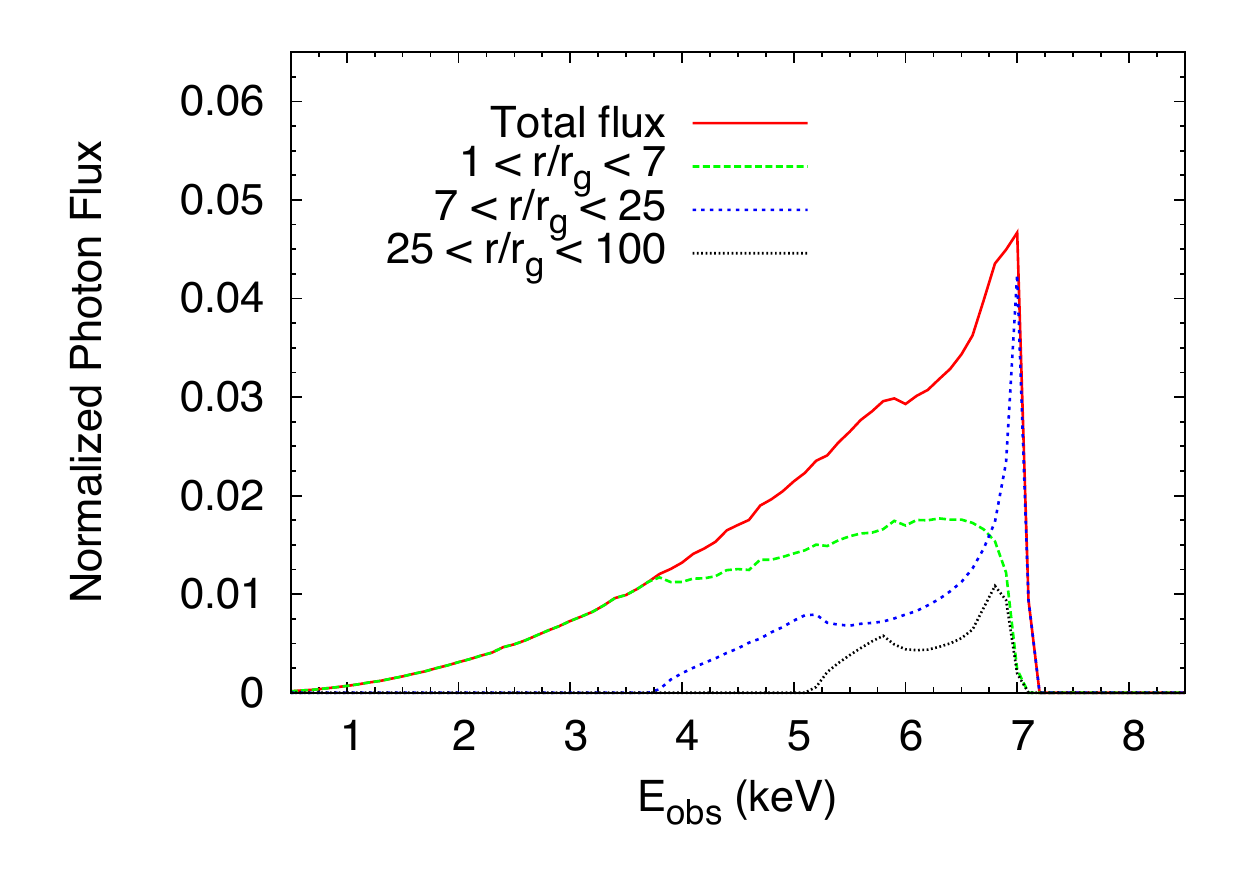}
\includegraphics[type=pdf,ext=.pdf,read=.pdf,width=7.5cm]{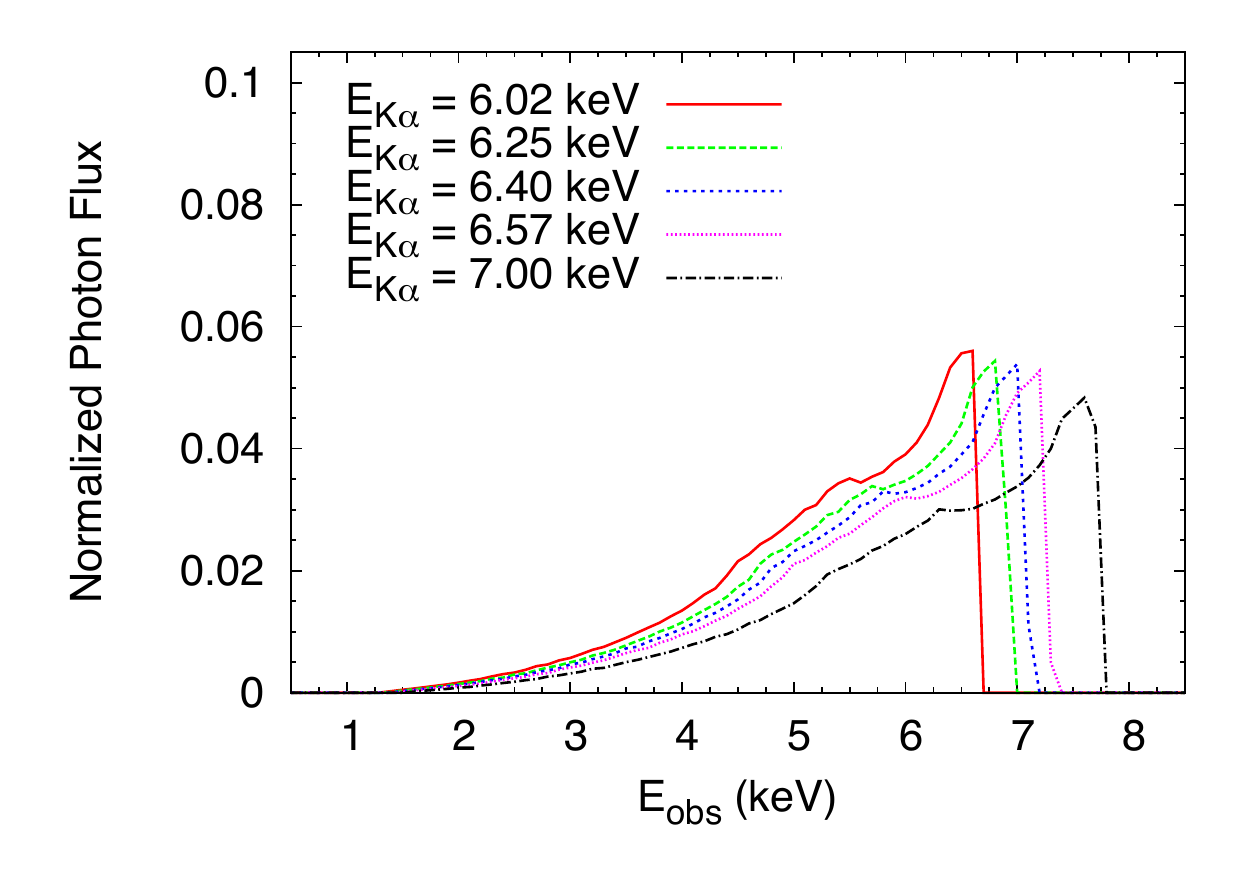}
\end{center}
\vspace{-0.5cm}
\caption{Left panel: K$\alpha$ iron line in Kerr spacetime with spin parameter 
$a_* = 1$, viewing angle $i = 45^\circ$, inner radius $r_{\rm in} = r_{\rm ISCO} = r_g$, 
outer radius $r_{\rm out} = 100 \, r_g$, index of the emissivity profile $q = 3$; 
the panel shows the total flux as well as the corresponding contributions from 
the disk regions $1 < r/r_g < 7$, $7 < r/r_g < 25$, and $25 < r/r_g < 100$. The high 
energy cut-off in the spectrum is due to Doppler blueshift at radii $7 < r/r_g < 25$.
Right panel: As in the right panel of Fig.~\ref{f1}, with viewing angle $i = 45^\circ$ 
and different values of $E_{\rm K \alpha}$. \label{f2}}
\end{figure}

The spin parameter $a_*$ determines the ISCO radius, ranging from $r_g$
for $a_* = 1$ (maximally rotating black hole and corotating accretion disk)
to $6 \, r_g$ for $a_* = 0$ (non-rotating black hole), and $9 \, r_g$ for $a_* = -1$
(maximally rotating black hole and counterrotating accretion disk). As shown in 
the left panel of Fig.~\ref{f1}, $a_*$ can be inferred from the low energy tail 
of the line, which is produced by the strong gravitational redshift at small
radii. The right panel of Fig.~\ref{f1} shows instead the effect of the inclination
angle $i$ on the line profile: a higher (lower) inclination angle moves the
high energy cut-off to higher (lower) energies, with small effects on the 
extension of the low energy tail\footnote{In Fig.~\ref{f1}, the photon flux is
normalized as $\int_{0}^{\infty} N(E) dE = 1$ (see Ref.~\cite{iron1}) and this 
causes a certain dependence of the low energy photon flux on $i$. While 
the inclination angle surely alters the Doppler redshift/blueshift, the 
effect is subdominant with respect to the gravitational redshift and therefore 
the low energy tail is mainly determined by the black hole spin parameter. 
If we focus on the low energy tail, we can use the normalization 
$\int_0^{3.5 \,\, {\rm keV}} N(E) dE = 1$. For a black hole with $a_* = 0.98$, the 
photon flux at 3.5~keV varies less than 8\% changing $i$ in the range 
$0^\circ - 65^\circ$, and something more in the range $0^\circ - 90^\circ$.}. Indeed, 
the cut-off is produced by Doppler blueshift at relatively large radii, approximately 
in the region $7 < r/r_g < 25$ (left panel of Fig.~\ref{f2}). Doppler boosting is 
maximum for edge-on disks ($i \rar 90^\circ$) and vanishes for face-on disks 
($i \rar 0^\circ$).

\begin{table*}
\begin{center}
\begin{tabular}{c c | c c c c |c}
\hline
\hline
{\small Source} & {\small$i$ (deg)} &  & {\small Iron} &  & & {\small $\Delta\alpha/\alpha$} \\
 && {\small $i$ range (deg)} & {\small $i$ (deg)} & {\small $a_*$} & {\small Ref.} &  \\
\hline
\hline
{\small Cygnus~X-1} & {\small $27.1\pm0.8$ (O)~\cite{orosz11}} & {\small free} & {\small $30\pm1$} & {\small $0.05\pm0.01$} & {\small \cite{miller09}} & {\small $0.005\pm0.003$} \\
 & $$ & {\small free} & {\small $32\pm2$} & {\small $0.88^{+0.07}_{-0.11}$} & {\small \cite{duro11}} & {\small $0.008\pm0.005$} \\ 
 & $$ & {\small free} & {\small $23.7^{+6.7}_{-5.4}$} & {\small $0.97^{+0.014}_{-0.02}$} & {\small \cite{fab12}} & {\small $-0.005^{+0.011}_{-0.007}$} \\ 
 & $$ & {\small free} & {\small $39.8^{+3.0}_{-4.2}$} & {\small $> 0.98$} & {\small \cite{fab12}} & {\small $0.023^{+0.008}_{-0.010}$} \\ 
 & $$ & {\small free} & {\small $> 40$} & {\small $> 0.83$} & {\small \cite{tom14}} & {\small $> 0.023$} \\
\hline
{\small GRS~1915+105} & {\small $66\pm2$ (J)~\cite{fen99}} & {\small $65 \le i \le 80$} & {\small $72\pm1$} & {\small $0.98\pm0.01$} & {\small \cite{mil13}} & {\small $0.027 \pm 0.014$} \\
\hline
{\small GRO~J1655-40} & {\small $70.2\pm1.2$ (O)~\cite{shafee}} & {\small $69 \le i \le 85$} & {\small $69^{+1}$} & {\small $0.98\pm0.01$} & {\small \cite{miller09}} & {\small $-0.006^{+0.010}$} \\
\hline
{\small 4U~1543-475} & {\small $20.7\pm1.5$ (O)~\cite{shafee}} & {\small $20 \le i \le 22$} & {\small $22_{-1}$} & {\small $0.3\pm0.1$} & {\small \cite{miller09}} & {\small $0.0016_{-0.0031}$} \\
 \hline
{\small XTE~J1550-564} & {\small $74.7\pm3.8$ (O)~\cite{orosz10}} & {\small $50 \le i \le 80$} & {\small $50^{+1}$} & {\small $0.76\pm0.01$} & {\small \cite{miller09}} & {\small $-0.023^{+0.004}$} \\ 
 &  & {\small $60 \le i \le 82$} & {\small $82_{-3}$} & {\small $0.55^{+0.15}_{-0.22}$} & {\small \cite{steiner}} & {\small $0.003_{-0.003}$} \\ 
  & {\small $71^{+12}_{-7}$ (J)~\cite{steiner12}} &  &  &  &  &  \\ 
\hline
\hline
\end{tabular}
\end{center}
\vspace{-0.2cm}
\caption{List of the stellar-mass black hole candidates for which the inclination 
angle $i$ can be inferred from the analysis of the K$\alpha$ iron line and from
at least another method. Strictly speaking, optical/near-infrared observations (O)
determine the inclination angle of the orbital plane of the binary system. The 
measurements of the inclination angle of the jet (J) should at least reflect the 
orientation of the black hole spin. See the text for details.}
\label{tab}
\end{table*}

\section{Constraints}

The position of the high energy cut-off, which is determined by $i$ in the
standard theory, may also be changed by a variation of the energy of the line. 
This is shown in the right panel of Fig.~\ref{f2}. The energy of the K$\alpha$ 
iron line is set by the Rydberg energy,
\be
E_{\rm K\alpha} \sim \frac{\alpha^2 m_e c^2}{2} (Z - 1)^2 \, ,
\ee
where $m_e$ is the electron mass and $Z = 26$ the atomic number of the 
iron. If we can have an independent measurement of the inclination angle of
the accretion disk, the determination of the position of the high 
energy cut-off of the K$\alpha$ iron line can potentially measure the value
of $\alpha^2 m_e$ in a gravitational potential relative to Earth of 
$\Delta\phi\approx 0.1$ (or the value of $\alpha$, if we assume that the 
electron mass cannot change). Let us note that here photons are emitted 
from the accretion disk, propagate in a Kerr spacetime, and eventually reach 
the X-ray detector, which measures the photon energy with the value of the 
fundamental constants on Earth. So, we are effectively comparing the energy
of the K$\alpha$ iron line in the region around the black hole with the one 
measured on Earth. We can potentially constrain the value
of fundamental constants with a single energy line because we can calculate
the redshift factor. In Refs.~\cite{webb1a,webb1b,webb2,berengut}, the authors have to
consider two lines with different $\alpha$-dependence because the redshift
factor is not known and therefore with a single line one can only measure
a combination of the redshift and of the value of $\alpha$.

Independent estimates of the inclination of the disk are possible. 
Detection of EEP violation relies on an ``apparent'' misalignment between the 
inclination angle of the inner disk (inferred from the iron line) and the angle of 
the orbital plane of the binary system (obtained from optical/near-infrared light curve
measurements) or the angle of the jet. The crucial assumption of this approach 
is therefore that there is no ``real'' misalignment between the inclination of the 
inner disk and the reference one. If the jet is powered by the black hole spin, strictly 
speaking the jet measurement provides the inclination angle of the spin. 
However, the Bardeen-Petterson effect should quickly align the inner accretion 
disk to the black hole spin~\cite{bp-effect}. If the jet is powered by the rotational 
energy of the disk, its measurement automatically gives the inclination angle 
of the inner disk. The orbital plane of the binary system is obtained from 
optical/near-infrared observations of the stellar companion and of the accretion 
disk at large scales. In the case of an initial misalignment with the black hole spin, the 
disk forces the black hole to alignment with a time scale of order $10^6 - 10^8$~yrs~\cite{maccarone,steiner12}. Such a time is typically shorter than the age of the system, 
so that (in the worst situation, i.e. measurement of the inclination angle of the orbital
plane and initial misalignment with the black hole spin) one could expect that 
only $1-2$\% of the black hole binary systems with an initial misalignment 
are still misaligned. Let us notice that the original 
alignment timescale estimated in Ref.~\cite{maccarone} was overestimated 
by a factor 50 due to a numerical error in Eq.~(6) of that paper.

Tab.~\ref{tab} shows the list of black holes in X-ray binary systems for which 
an estimate of the inclination angle from both dynamical/jet methods and analysis 
of the iron line are reported in the 
literature~\cite{orosz11,fen99,shafee,orosz10,steiner,miller09,duro11,fab12,tom14,mil13}. 
The first column reports the name of the source and the second column 
the estimate of $i$ from optical/near-infrared light curve (O) and jet (J) 
measurements. The third, fourth, fifth, and sixth 
columns are for the measurements obtained from the analysis of 
the K$\alpha$ iron line. As the shape of the line depends on several free 
parameters, the fitting procedure is very time consuming, while the data 
are not always very good. It is thus common to restrict the range of the free 
parameters on the basis of independent estimates. The fourth column 
shows exactly the range of $i$ adopted in these studies. Cygnus~X-1 has 
been studied in Refs.~\cite{miller09,duro11,fab12,tom14}, always
leaving $i$ completely free. The low spin parameter found in Ref.~\cite{miller09} 
seems to be due to both improper data state (the source was not in the 
high/soft state) and the improper usage in the continuum model in 
extracting the skewed iron line profile (see discussion in Section~7.1 
of~\cite{gou}). The two inclination angles found in Ref.~\cite{fab12}
refer to two different models: the lower value is provided by a model with 
two emissivity indices obtained from the fit, while the higher one by a
model with $q=3$. The authors of Ref.~\cite{tom14} consider several
models, and the ones that provide good fits have inclination angles
ranging between $40^\circ$ and $69^\circ$. Such a model-dependent
inclination angle in the results of Ref.~\cite{tom14} may come from missing
physics, which was not important in observations of satellites like 
Chandra and XMM-Newton, but that cannot be ignored for NuSTAR.
The seventh column of Tab.~\ref{tab} shows the measurements of the
variation of the fine structure constant $\alpha$ (assuming that $m_e$ is
a true constant) obtained by comparing a line at $E = 6.40$~keV with
$i$ given by the fifth column with lines resulting from a variable rest-frame 
energy and an inclination angle given by the first column of Tab.~\ref{tab}.
Since the aim of the present paper is just to propose this technique to
constrain possible variation of the fine structure constant in strong gravitational
field, the bounds on $\Delta\alpha/\alpha$ reported in Tab.~\ref{tab} have 
been obtained by comparing two theoretical models, not from actual data,
which simplifies the analysis a lot.
The other model parameters are fixed, because the high energy cut-off
of the iron line profile is not very sensitive to them (see Appendix~\ref{app-c}
for the details). The procedure to determine $\Delta\alpha/\alpha$ 
is described in Appendix~\ref{app-c}.

\section{Discussion}

Based on the simplified approach discussed in Appendix~\ref{app-c} that does 
not analyze real data but compares two theoretical models with the measurements 
reported in the literature, one can find the constraints on possible variations of the 
fine structure constant $\alpha$ in the gravitational fields of stellar-mass black 
holes shown in the last column of Tab.~\ref{tab}. The uncertainties on 
$\Delta \alpha/\alpha$ are smaller than the differences between the bounds inferred
from different measurements and sources, which suggests that there are systematics
effects not properly taken into account. However, if we consider the spread of these
constraints as uncertainty, we can argue that the X-ray data are consistent with
no variations of $\alpha$ larger than some percent. If we exclude the measurements
reported in Ref.~\cite{tom14}, all the constraints converge to $|\Delta\alpha/\alpha| < 3$\%.
In Ref.~\cite{tom14}, the authors consider different models and the ones with
a good fit require an inclination angle in the range $40^\circ$ and $69^\circ$.
If the dynamical estimate of $i$ from Ref.~\cite{orosz11} is correct and there is not
a misalignment between the inner accretion disk and the orbital plane of the binary
system, these measurements would suggest a non-vanishing 
$\Delta\alpha/\alpha$ in the range $2-10$\%. However, a $\Delta\alpha/\alpha$ 
exceeding 3\% would disagree with all the other results reported in the literature
and can be probably excluded.

All the bounds have to be taken with some caution. In the case of the measurements
reported in Ref.~\cite{fab12} for the inclination angle of the inner disk of Cygnus~X-1,
the author find $i = 23.7^\circ$ consistent with the dynamical method for a model
with two emissivity indices, $q_1$ and $q_2$, both assumed as free parameters
and determined in the fitting procedure. The angle $i=39.8^\circ$ is instead found
for a model with $q=3$. While the first model should be more reliable, this
example shows the impact on the choice of the exact/wrong model to constrain
variations of the fine structure constant.

In the case of the black hole binaries in 4U~1543-475, GRO~J1655-40, and
XTE~J1550-564, the inclination angle is not really a free parameter, but it
is allowed to vary in a small range. For 4U~1543-475 studied in~\cite{miller09}, 
the range is so small that $i$ is effectively constrained by the optical observations, 
and therefore the bound on $\alpha$ is useless. For GRO~J1655-40 and 
XTE~J1550-564, the initial range of $i$ is larger, but the fits require values 
at the boundary of the allowed range, so meaningful constraints can be put 
only on one side. For XTE~J1550-564, the two measurements of the iron line
are not consistent
each other, and there exists also an estimate of the inclination angle of the
jet whose uncertainty exceeds the allowed range of $i$ in the analysis of 
Refs.~\cite{miller09,steiner12} (and for this reason it has not been used in 
Tab.~\ref{tab}).

As already pointed out, the bounds reported in Tab.~\ref{tab} have been obtained
from a simple analysis without using actual X-ray data, just to show how this
approach can work and get a crude estimate of possible constraints. 
In the case of real data, the situation is more complicated.
For the highly ionized gas in X-ray binaries, atoms are in different ionization states 
and therefore the spectrum is a combination of several lines, ranging from 6.6 to 
7.0~keV~\cite{ka-rev1,ka-rev2}. The exact spectrum depends on the ``ionization 
parameter'', which changes the relative contribution of the different ionization states. 
Different groups have computed tables of reflection spectra via radiative transfer 
for an illuminated atmosphere for a range of metallicities, densities, spectral indices, 
incident fluxes, and ionization parameters. In the studies reported in Tab.~\ref{tab}, 
the authors use these tables, not the value 6.4~keV. The fact that there is a spectrum 
rather than a single line makes the analysis more complicated, but it should not 
alter the ability to constrain $\Delta\alpha/\alpha$ if all the astrophysical details are 
properly taken into account. The ionization parameter is inferred by the fit, but 
the key-point is that the correlation between the ionization parameter and the 
inclination angle is not large~\cite{priv}. If the ionization parameter is 
determined correctly, the exact spectrum is known. Let us also note that even the
cleanest neutral 6.4~keV line is never a single line, but a blend of two lines 
that differ in energy by about 0.3\%. That is further complicated by the fact that
multiple ionization states each with multiple iron ion species radiate in different
regions of the disk. The determination of the exact spectrum is currently the main 
source of uncertainty for the estimate of the viewing angle which, otherwise, 
could be determined with higher precision. While it is beyond the scope of the
present paper to properly address this problem, it is definitively a key-issue in
the future use of the K$\alpha$ iron line to test the value of the fine structure 
constant in the strong gravitational field of black holes. Intuitively, it seems that
this is just a complication, but that it does not affect the method once the exact
spectrum is determined, because it should just manifest as a shift which
cancels out in the fit. However, the issue requires a much more detailed investigation,
to exclude a possible correlation between the value of $\alpha$ and the estimate
of the spectrum, which otherwise would invalidate or at least weaken this
method to constrain $\alpha$ from the iron line profile.

In the study reported in this paper, the Kerr solution has been assumed as
background metric. Actually, black holes in alternative theories of gravity may 
be described by other solutions. In those theories in which modifications to 
general relativity are Planck scale suppressed in the infrared, the Kerr metric 
is a good approximation. In the other cases, the position of the high energy 
cut-off of the iron line may be slightly different from the one around a Kerr black 
hole with the same viewing angle. That is mainly the result of a different orbital 
velocity of the gas. The angle $i$ inferred from the iron line assuming (erroneously) 
the Kerr metric can be the correct one or be different by 1-2 degrees with respect 
to the true angle (sometimes even more), depending on the specific 
background~\cite{iron1,iron2,iron4}. An observation confirming the same value 
of $\alpha$ as the one on Earth could suggest that this is not the case, but in 
principle one cannot really exclude a fine compensation between a non-vanishing 
variation of $\alpha$ and a different Doppler blueshift producing an effect of the 
same magnitude but of opposite sign. Such a possibility can be ruled out in the 
case of measurements of $i$ from sources with different inclination angle.

Since the value of fundamental constants like $\alpha$ may be determined by 
the vacuum expectation value of some fields, the analysis of the K$\alpha$ iron 
line can test the existence of hairy black holes~\cite{hbh1,hbh2,hbh3,hbh4,hbh5}.
The presence of fields non-universally coupled to matter would affect also 
the motion of particles. In other words, we should expect also a violation of the 
WEP. However, one can see that such a possibility does not affect the calculations 
of the K$\alpha$ iron line. As shown in Appendix~\ref{app-b}, photons still
follow the null geodesics of the spacetime, and therefore the calculation of the
photon trajectories from the disk to the plane image of the distant observer
is unaltered. Modifications of the photon trajectories would require additional
new physics, like a term in the Lagrangian that provides mass to the photons.
The properties of the trajectories of the gas's particles in the disk are also
not affected by possible variations of $\alpha$, because they follow
circular orbits, and therefore they feel the same value of $\alpha$
as a consequence of the axisymmetric background.

\section{Summary and conclusions}

Time and/or space variations of fundamental constants violate the EEP.
The validity of the EEP is a fundamental ingredient in general relativity, but 
its violation is often expected in alternative theories of gravity and in 
theories aiming at unifying gravity with the forces of the Standard Model. 
Today there is no clear evidence of variations of fundamental constants, 
but observations of quasar spectra at high redshift hint at different values 
of the fine structure constant $\alpha = e^2/\hbar c$ with respect to the one 
measured on Earth~\cite{webb1a,webb1b,webb2}. A possible dependence 
of $\alpha$ on gravitational fields has been recently constrained in 
Ref.~\cite{berengut} from the far-UV spectra of a white-dwarf star, thus probing 
a gravitational potential relative to Earth of $\Delta\phi \approx 5\cdot 10^{-5}$, 
and the bound is $|\Delta \alpha/\alpha| \lesssim 10^{-4}$. In this work, I showed 
that the position of the high energy cut-off in the profile of broad K$\alpha$ iron 
lines observed in the X-ray spectrum of black holes can be used to test $\alpha$ 
in much stronger gravitational fields, where $\Delta\phi \approx 0.1$, when an 
independent measurement of the disk's inclination angle is available. At present,
there are systematic effects not fully under control, and therefore this technique
cannot provide robust and stringent constraints on $\alpha$. Based on a
simplified approach that does not analyze real data but compares two theoretical
models with the measurements reported in the literature, I obtained the bounds
in the last column of Tab.~\ref{tab}. The uncertainty on these constraints is smaller
than their difference, which suggests that at least in some studies the systematics
has not been properly taken into account. However, all the measurements are
consistent with no variation of $\alpha$ large than some percent. This can be
considered the present crude bound on possible variations of $\alpha$ 
with this approach. Even if we 
consider the measurements reported in Ref.~\cite{tom14}, in which the 
authors find an inclination angle exceeding $40^\circ$ and up to $69^\circ$,
the data could still be explained with a variation of $\alpha$ lower than $2-10$\%
(but a $\Delta \alpha/\alpha > 3\%$ seems unlikely because in disagreement 
with all the other studies). 
If in the future it is possible to have all the systematics under control,
constraints of $\Delta \alpha/\alpha$ better than 1\% with this technique do not 
seem to be out of reach.

\appendix

\section{Theories violating the Einstein Equivalence Principle \label{app-a}}

The theories satisfying the EEP are called ``metric theories of gravity'' as they
meet the following assumptions (for more details, see Ref.~\cite{will}):
\begin{enumerate}
\item The spacetime is endowed with a symmetric metric.
\item The trajectories of freely-falling test-particles are the geodesics of 
that metric.
\item In any local freely-falling reference frame, the non-gravitational laws of 
physics reduce to the ones of special relativity.
\end{enumerate}
General relativity is clearly a metric theory of gravity. Tensor-scalar theories
are metric theories of gravity if the additional degrees of freedom universally 
couple to matter. For instance, if we consider a theory in which the gravity
sector is described by the spacetime metric $g_{\mu\nu}$ and by a scalar
field $\phi$, and gravity is universally coupled to matter, the total action looks 
like 
\be\label{eq-act}
S &=& S_{\rm g} [g_{\mu\nu}, \phi]
+ S_{\rm m} [\psi_{\rm m1}, \psi_{\rm m2}, ..., A^2(\phi)g_{\mu\nu}] \, .
\ee
The crucial point that determines the fact that the action in~(\ref{eq-act})
belongs to a metric theory of gravity is that the matter sector responds to the 
metric 
\be\label{eq-conf}
\tilde{g}_{\mu\nu} = A^2(\phi)g_{\mu\nu} \, ,
\ee
and therefore we can perform the conformal transformation above and find
that all the particles follow the geodesics of the metric $\tilde{g}_{\mu\nu}$
and that the result of any local non-gravitational experiment must meet both 
the LLI and the LPI. On the other hand, if the new degrees of freedom do not
universally couple to matter, such a transformation is not possible. If the total
action is
\be
S &=& S_{\rm g} [g_{\mu\nu}, \phi]
+ S_{\rm m1} [\psi_{\rm m1}, A^2(\phi)g_{\mu\nu}]
+ S_{\rm m2} [\psi_{\rm m2}, B^2(\phi)g_{\mu\nu}] \, ,
\ee
we can still perform the conformal transformation in Eq.~(\ref{eq-conf}),
but now we find that the particles associated to the field $\psi_{\rm m1}$
follow the geodesics of the metric $\tilde{g}_{\mu\nu}$, but the ones associated
to the field $\psi_{\rm m2}$ do not; that is, the WEP is violated. The transformation
\be
\hat{g}_{\mu\nu} = B^2(\phi)g_{\mu\nu}
\ee
does not fix the problem: now the particles associated to $\psi_{\rm m2}$ follow
the geodesics of the metric $\hat{g}_{\mu\nu}$, but the ones associated to 
$\psi_{\rm m1}$ do not. The EEP turns out to be violated in many extensions of 
general relativity.

Let us now assume that all the elementary particles responds to the metric $g_{\mu\nu}$.
The electromagnetic action can be~\cite{dam1,dam2,dam3}
\be\label{eq-em}
S_{\rm em} = \frac{1}{4 g_{\rm em}^2} \int \sqrt{-g} \, C^2(\phi) \, g^{\mu\nu} \, 
g^{\rho\sigma} \, F_{\mu\rho} \, F_{\nu\sigma} \, d^4 x \, ,
\ee
where $g_{\rm em}$ is the ``actual'' electromagnetic coupling and $F_{\mu\nu}$ 
is the strength of the electromagnetic field. The ``effective'' electromagnetic 
constant turns out to be 
\be\label{eq-alpha}
\alpha = g_{\rm em}^2 C^{-2}(\phi) \, ,
\ee 
and, if $\phi$ is not constant, the LPI can be violated. From the new
Maxwell's equations, one can see that photons still follow the null geodesics
of the metric $g_{\mu\nu}$ (see Appendix~\ref{app-b}). In general, a space
and/or time variations of the fine structure constant implies also a violation of 
the WEP, because the mass of particles like protons and neutrons receives 
contributions from the electromagnetic interaction and therefore spacetime 
variations of $\alpha$ necessarily imply spacetime variations of their 
masses, which violates the equivalence between inertial and gravitational mass. 
However, the calculations of the profile of the K$\alpha$ iron line only involve
the properties of equatorial circular orbits of the gas's particles, which are 
not affected by a possible dependence of $\alpha$ on $\phi$ for axisymmetric 
backgrounds.

\section{Photon propagation in the background metric \label{app-b}}

In the standard theory, Maxwell's equations $\nabla_\nu F^{\mu\nu} = 0$
imply that photons follow the null geodesics of the background metric,
see e.g. Section~22.5 of Ref.~\cite{book-mtw}. Indeed, from the identity
\be
\nabla_\nu \nabla^\mu A^\nu = 
\nabla^\mu \nabla_\nu A^\nu + R^\mu_{\;\; \nu} A^\nu
\ee
and the Lorentz gauge condition $\nabla_\nu A^\nu = 0$, Maxwell's
equations $\nabla_\nu F^{\mu\nu} = 0$ become
\be\label{eq-m}
R^\mu_{\;\; \nu} A^\nu - \nabla_\nu \nabla^\nu A^\mu = 0 \, .
\ee
In the geometrical optics approximation, the electromagnetic potential 
$A_\mu$ can be written as
\be\label{eq-a}
A_\mu = {\rm Re} \left[ \left(a_\mu + \epsilon b_\mu 
+ \epsilon^2 c_\mu + . . . \right) e^{i \Theta/\epsilon}\right] \, ,
\ee
where $\epsilon \ll 1$ is an expansion parameter. Light rays are 
defined as the curves normal to the surfaces of constant phase $\Theta$.
The wave vector $k^\mu$ is given by $k^\mu = \nabla^\mu \Theta$.
If we plug Eq.~(\ref{eq-a}) into Eq.~(\ref{eq-m}), we have
\be
\hspace{-1cm}
&&{\rm Re} \Big\{ \Big[ \frac{1}{\epsilon^2} k_\nu k^\nu 
\left(a^\mu + \epsilon b^\mu + \epsilon^2 c^\mu + . . . \right)
+ \nonumber\\ &&\hspace{1cm}
- \frac{2i}{\epsilon} k^\nu \nabla_\nu
\left(a^\mu + \epsilon b^\mu + \epsilon^2 c^\mu + . . . \right)
- \frac{2i}{\epsilon} \left( \nabla_\nu k^\nu \right)
\left(a^\mu + \epsilon b^\mu + \epsilon^2 c^\mu + . . . \right)
+ \nonumber\\ &&\hspace{1cm} + \nabla_\nu \nabla^\nu
\left(a^\mu + \epsilon b^\mu + \epsilon^2 c^\mu + . . . \right)
+ R^\mu_{\;\; \nu}
\left(a^\nu + \epsilon b^\nu + \epsilon^2 c^\nu + . . . \right)
\Big] e^{i \Theta/\epsilon}\Big\} = 0 \, .
\ee
The leading order in the $\epsilon$-expansion is $O(\epsilon^{-2})$
and gives $k_\nu k^\nu a^\mu = 0$, i.e. $k^\nu$ is a null vector.
Since $\nabla_\nu k_\mu = \nabla_\mu k_\nu$, we have
\be
\nabla_\mu \left(k_\nu k^\nu \right) = 0 \;\;\; \Rightarrow \;\;\;
k^\nu \nabla_\mu k_\nu = 0 \, ,
\ee
which implies that photon trajectories are the null geodesics of the 
metric $g_{\mu\nu}$.

If the action of the electromagnetic sector is the one in Eq.~(\ref{eq-em}),
the variation of the electromagnetic potential $A_\mu$ provides
modified Maxwell's equations
\be
\nabla_\nu \left[ C^2(\phi) F^{\mu\nu}\right] = 0 \, .
\ee
If we use Eq.~(\ref{eq-a}), we have
\be
R^\mu_{\;\; \nu} A^\nu - \nabla_\nu \nabla^\nu A^\mu +
\frac{2}{C} \left( \partial_\nu C \right)
\left(\nabla^\mu A^\nu - \nabla^\nu A^\mu\right) = 0 \, .
\ee
Since $\nabla^\mu A^\nu - \nabla^\nu A^\mu = O(\epsilon^{-1})$, the leading order
term of Maxwell's equations remains $k_\nu k^\nu a^\mu = 0$ and photons still 
follow the null geodesics of the background metric.

\begin{figure}
\begin{center}
\includegraphics[type=pdf,ext=.pdf,read=.pdf,width=7.5cm]{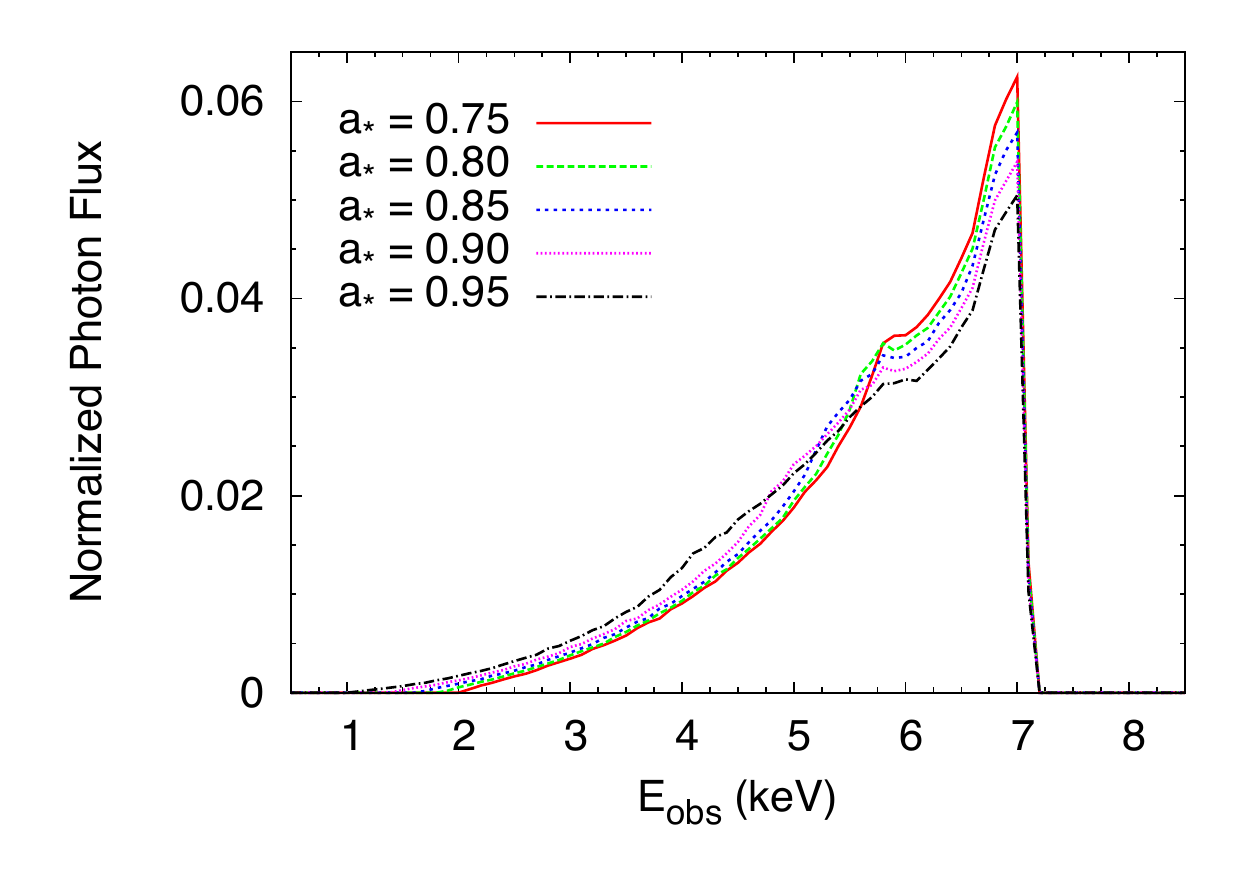}
\includegraphics[type=pdf,ext=.pdf,read=.pdf,width=7.5cm]{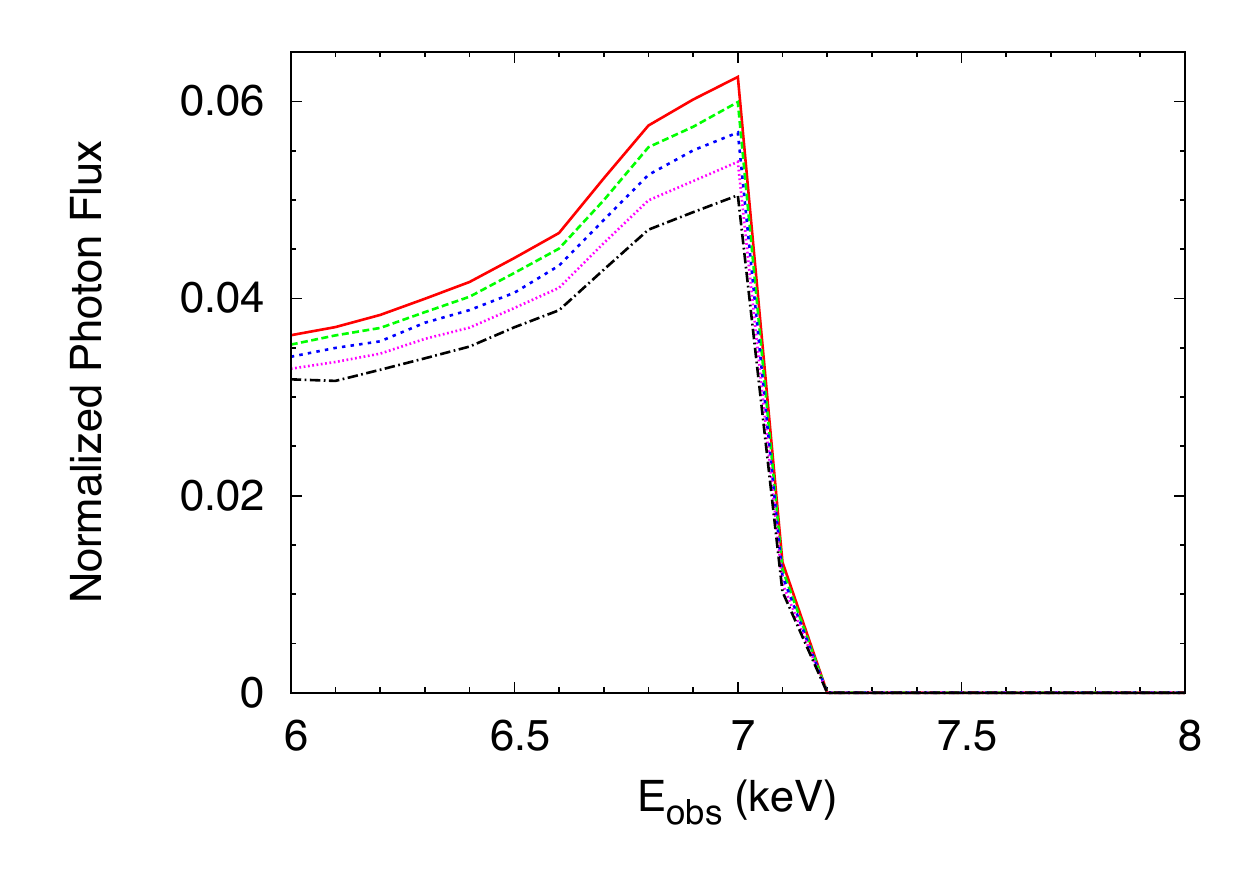}
\end{center}
\vspace{-0.5cm}
\caption{Iron line profile for different values of the spin parameter $a_*$ (the right 
panel is just the enlargement of the left one in the energy range 6-8 keV). The
other parameters are: viewing angle $i = 45^\circ$, inner radius 
$r_{\rm in} = r_{\rm ISCO}$, outer radius $r_{\rm out} = r_{\rm ISCO} + 100 \, r_g$,
index of emissivity profile $q = 3$.}
\label{f1a}
\vspace{1.0cm}
\begin{center}
\includegraphics[type=pdf,ext=.pdf,read=.pdf,width=7.5cm]{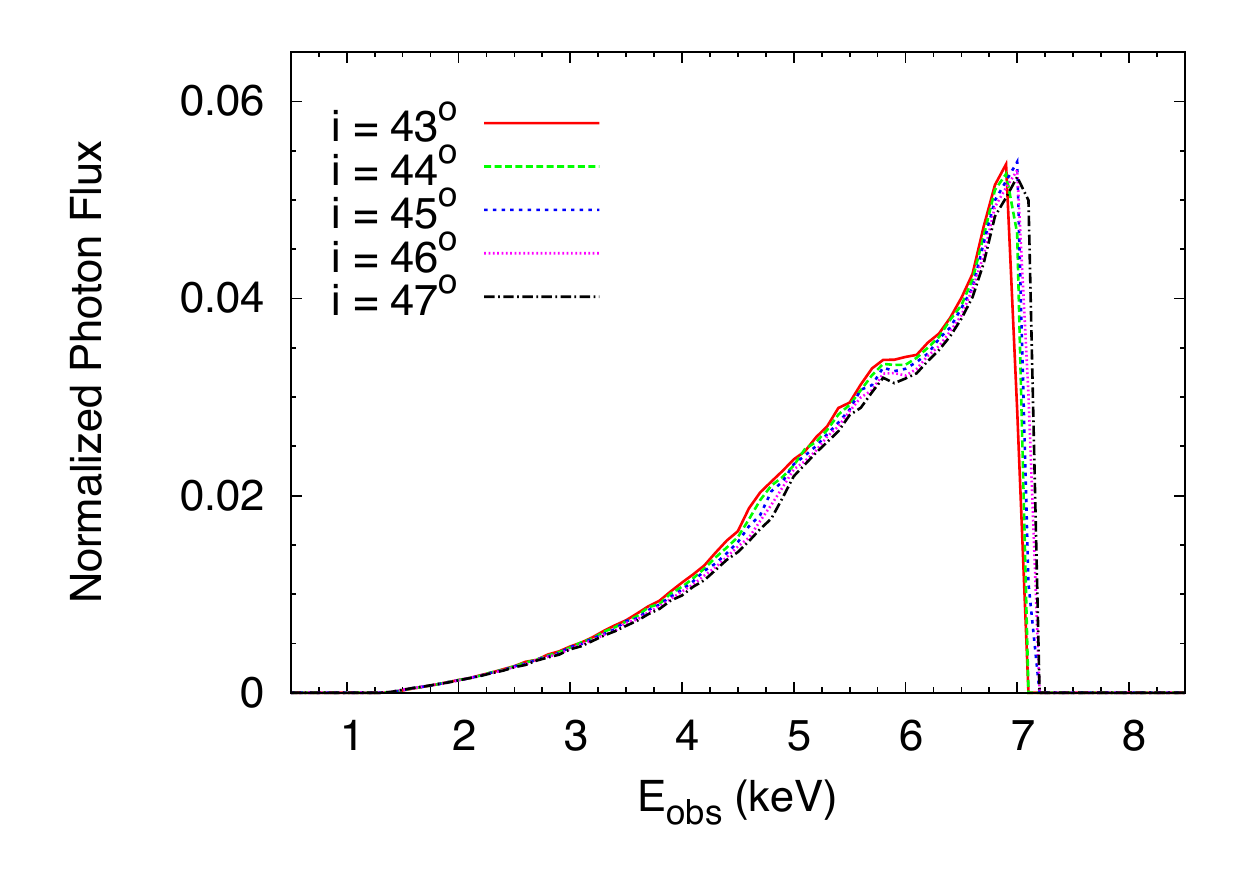}
\includegraphics[type=pdf,ext=.pdf,read=.pdf,width=7.5cm]{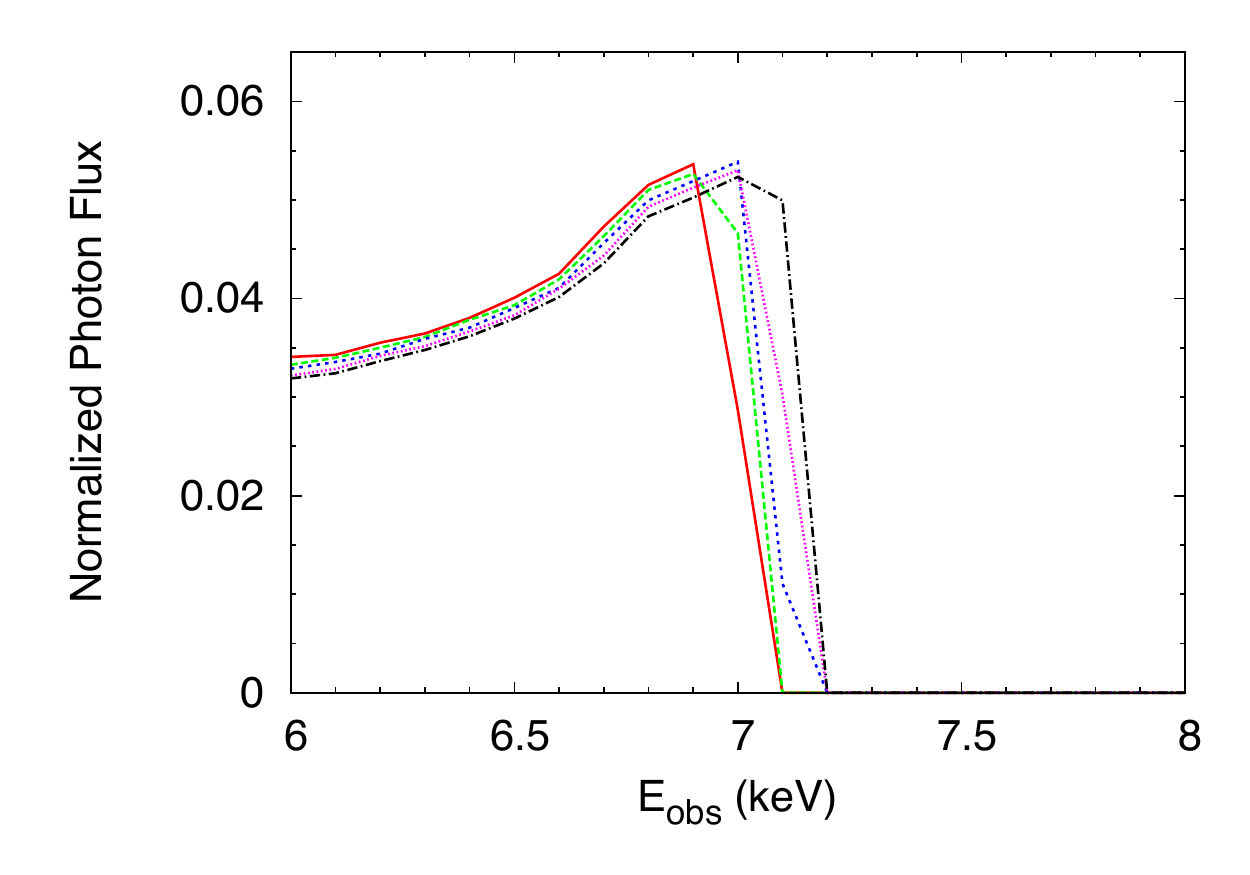}
\end{center}
\vspace{-0.5cm}
\caption{Iron line profile for different values of the viewing angle $i$ (the right 
panel is just the enlargement of the left one in the energy range 6-8 keV). The
other parameters are: spin parameter $a_* = 0.9$, inner radius 
$r_{\rm in} = r_{\rm ISCO}$, outer radius $r_{\rm out} = r_{\rm ISCO} + 100 \, r_g$,
index of emissivity profile $q = 3$.}
\label{f2a}
\end{figure}

\begin{figure}
\begin{center}
\includegraphics[type=pdf,ext=.pdf,read=.pdf,width=7.5cm]{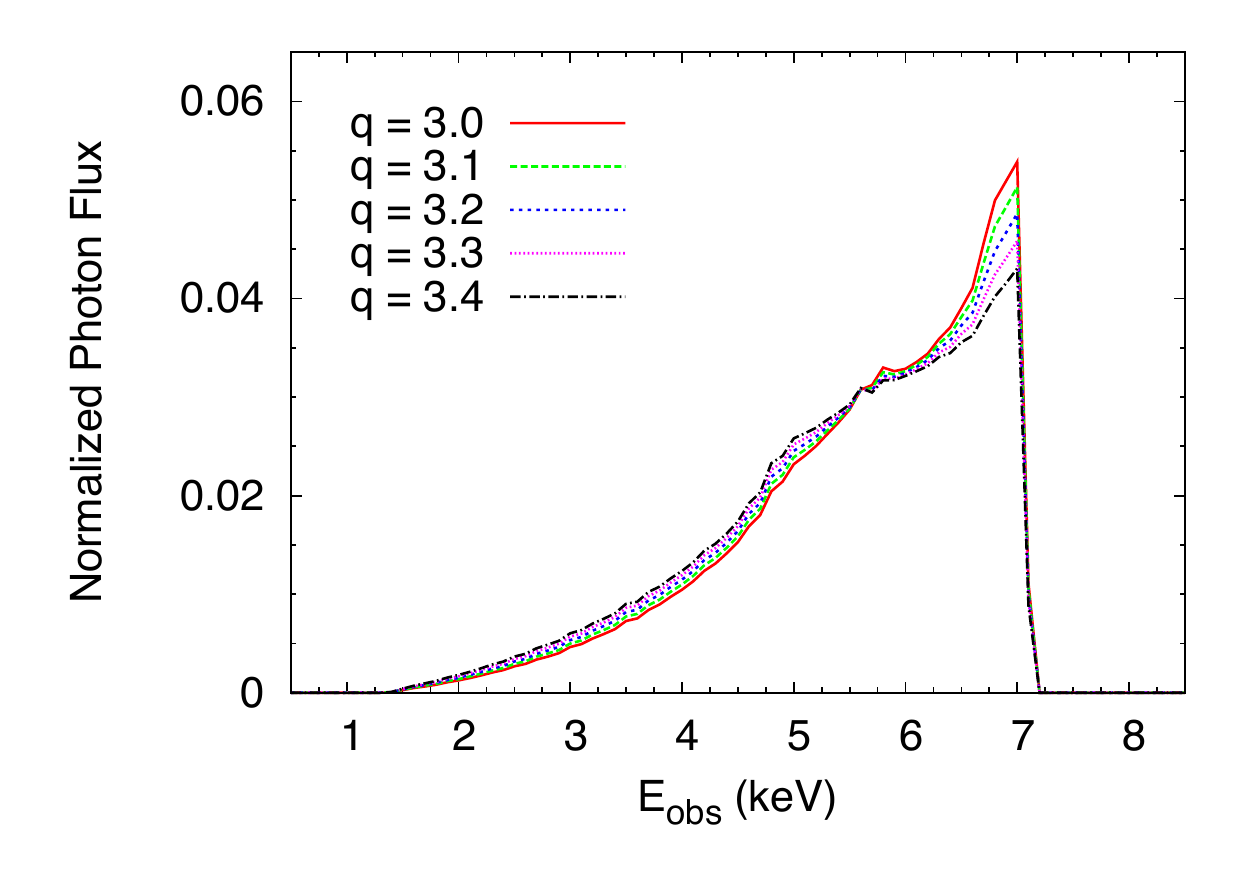}
\includegraphics[type=pdf,ext=.pdf,read=.pdf,width=7.5cm]{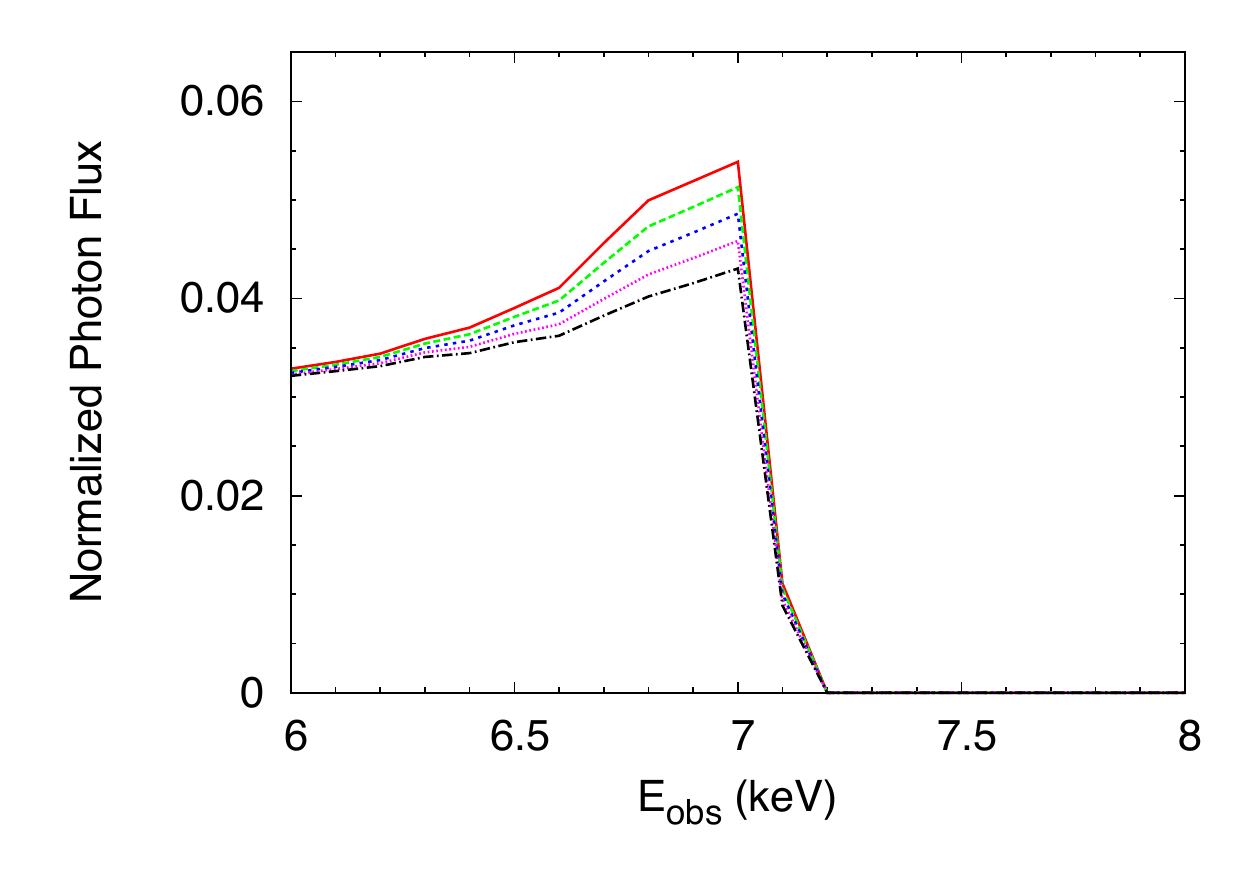}
\end{center}
\vspace{-0.5cm}
\caption{Iron line profile for different values of the index of emissivity profile 
$q$ (the right panel is just the enlargement of the left one in the energy range 
6-8 keV). The other parameters are: spin parameter $a_* = 0.9$, viewing 
angle $i = 45^\circ$, inner radius $r_{\rm in} = r_{\rm ISCO}$, outer radius 
$r_{\rm out} = r_{\rm ISCO} + 100 \, r_g$.}
\label{f3a}
\vspace{1.0cm}
\begin{center}
\includegraphics[type=pdf,ext=.pdf,read=.pdf,width=7.5cm]{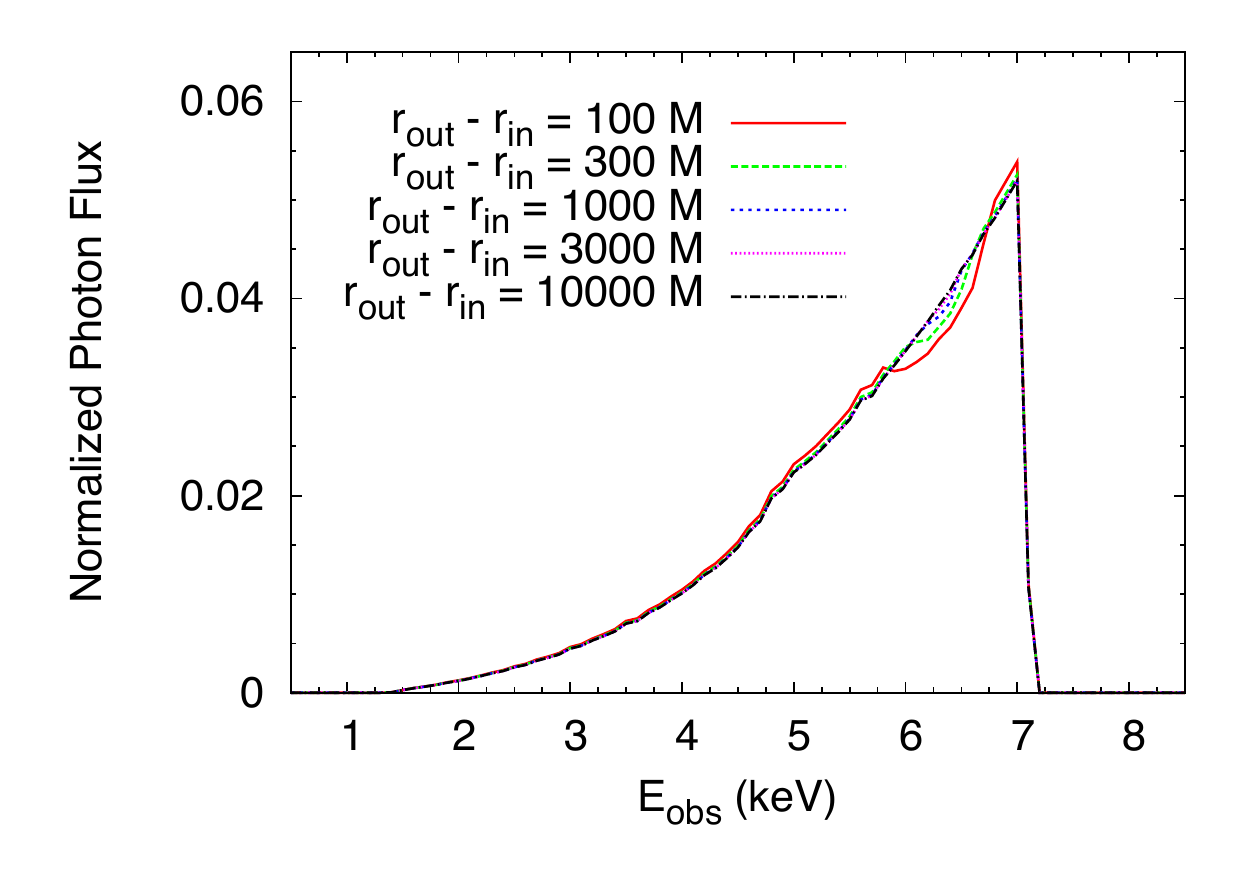}
\includegraphics[type=pdf,ext=.pdf,read=.pdf,width=7.5cm]{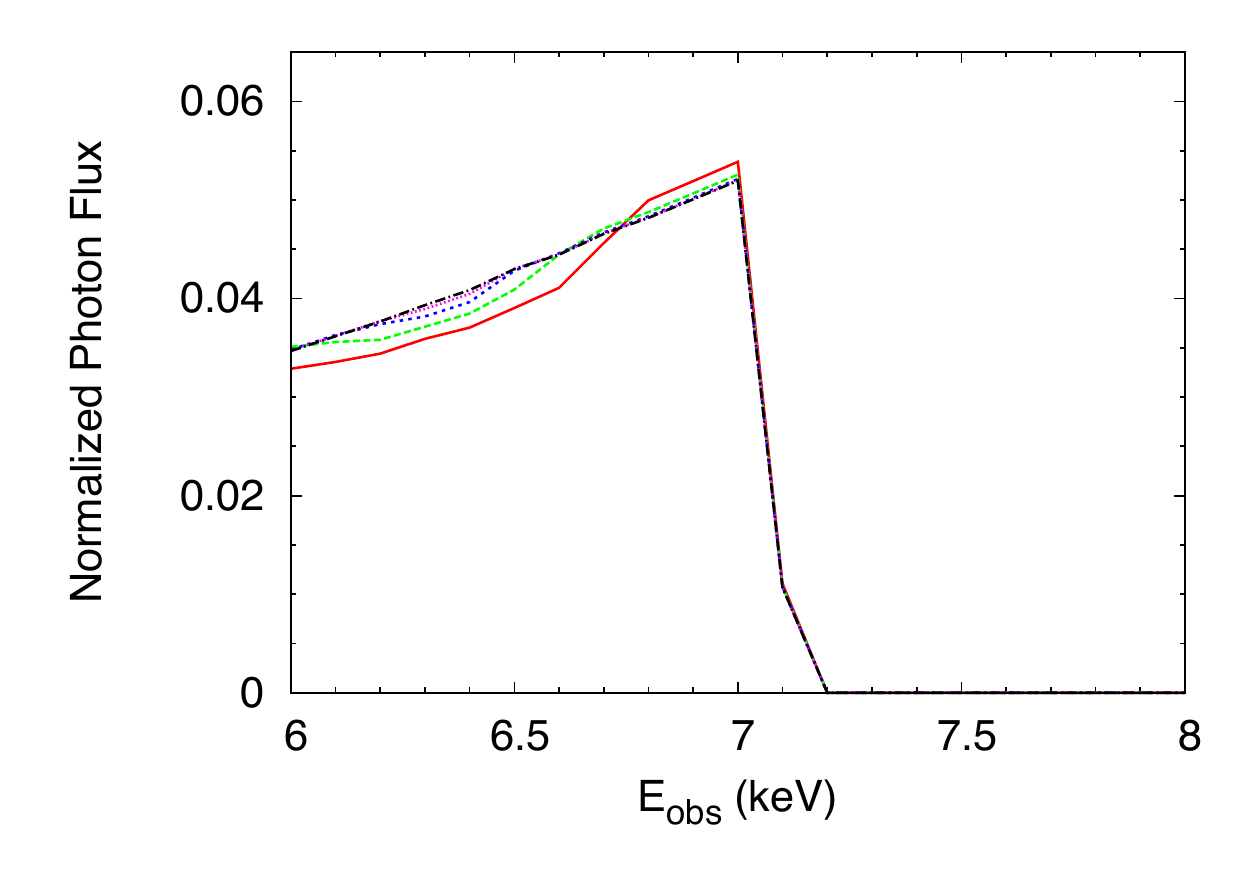}
\end{center}
\vspace{-0.5cm}
\caption{Iron line profile for different values of the outer radius $r_{\rm out}$ 
(the right panel is just the enlargement of the left one in the energy range 6-8 
keV). The other parameters are: spin parameter $a_* = 0.9$, viewing angle 
$i = 45^\circ$, inner radius $r_{\rm in} = r_{\rm ISCO}$, index of emissivity 
profile $q = 3$.}
\label{f4a}
\end{figure}

\section{Error analysis \label{app-c}}

Figs.~3-6 show the effect of the model parameters (spin $a_*$, viewing angle
$i$, outer radius $r_{\rm out}$, index of emissivity profile $q$) on the high energy
cut-off of the iron line profile. Uncertainties like $\Delta a_* \sim 0.05 - 0.10$,
$\Delta i \sim 1^\circ - 2^\circ$, and $\Delta q \sim 0.1 - 0.2$ are typical values found
in the iron line analysis. {\it The plots show that the effect of the viewing angle is 
qualitatively different from the one of the other parameters}. 
Such a statement can be quantitatively justified with the error analysis described 
below. For a crude estimate aiming at showing the basic idea of the proposal of
the present paper, we can restrict the attention only to the viewing angle when 
we constrain possible variations of the fine structure constant.

The value of the K$\alpha$ iron line in the strong gravitational field of a black hole
can be estimated by comparing the iron line profile with $E = 6.4$~keV and the angle 
inferred by the iron line fit (column 5 in Table~I) with the profiles calculated with the
``actual'' energy $E$ (which may be different from 6.4~keV because of a different
value of $\alpha$) and the angle obtained from dynamical/jet measurements (column 2
in Table~I). The latter are clearly 
independent of the values of fundamental constants close to the compact object.
If the other parameter are fixed, the reduced $\chi^2$ is
\be\label{eq-1}
\chi^2_{\rm red} (E_{\rm K\alpha})
&=& \frac{\chi^2}{n - p} = \nonumber\\
&=&\frac{1}{n-p} \sum_{i = 1}^{n} \frac{\left[
N_i (E_{\rm K\alpha},\tilde{a}_*,\tilde{i},\tilde{q},\tilde{r}_{\rm out})
- N_i (\tilde{E}_{\rm K\alpha},\tilde{a}_*,\tilde{i}_{\rm iron},\tilde{q},\tilde{r}_{\rm out}) 
\right]^2}{\sigma^2_i} \, ,
\ee
where the summation is performed over $n$ sampling energies $E_i$, $p$ is
the number of free parameters in the model being fitted (here $p=1$), 
and $N_i$ is the normalized photon fluxes in the energy bin $[E_i,E_i+\Delta E]$.
$\tilde{a}_*$, $\tilde{i}_{\rm iron}$, $\tilde{q}$, and $\tilde{r}_{\rm out}$ 
have the values measured by the iron line fit, while $\tilde{i}$ has the 
one inferred by the optical/near-infrared light curve observations or jet measurements. 
$\tilde{E}_{\rm K\alpha} = 6.4$~keV.
The resulting energy $E_{\rm K\alpha}$ is the one that minimizes the reduced 
$\chi^2$. This provides the central value of $\Delta \alpha/\alpha$ in the last column
of Table~I in the paper. The $\Delta E$ used in these calculations is 100~keV,
which is roughly the resolution of current X-ray facilities, and the summation is
performed over a small energy range of 2~keV around the peak of the line.
While data are often more sensitive to lower energies, depending on the detector 
and its effective area, because of the reduced energy range here the same 
weighting has been assigned to different energies.

The uncertainty on $\Delta \alpha/\alpha$ reported in Tab.~I is obtained by
replacing the 1-$\sigma$ values of $\tilde{i}$ and $\tilde{i}_{\rm iron}$
in Eq.~\eqref{eq-1}. If we compute
\be
\chi^2_{\rm red} (E_{\rm K\alpha}) =
\frac{1}{n-p} \sum_{i = 1}^{n} \frac{\left[
N_i (E_{\rm K\alpha},\tilde{a}_*,\tilde{i} 
\pm \Delta \tilde{i},\tilde{q},\tilde{r}_{\rm out})
- N_i (\tilde{E}_{\rm K\alpha},\tilde{a}_*,\tilde{i}_{\rm iron} 
\pm \Delta \tilde{i}_{\rm iron},\tilde{q},\tilde{r}_{\rm out}) 
\right]^2}{\sigma^2_i} \, ,
\nonumber\\
\ee
we get four different energies (coming from the four different combinations of
$\tilde{i} \pm \Delta \tilde{i}$ and $\tilde{i}_{\rm iron} 
\pm \Delta \tilde{i}_{\rm iron}$). We can then take the two extreme values as
upper and lower values of $\Delta \alpha/\alpha$. As we are
considering only the central value for the other model parameters ($\tilde{a}_*$, 
$\tilde{q}$, and $\tilde{r}_{\rm out}$), neglecting their uncertainty, it is not really 
important the energy range under consideration, in the sense that the final 
result does not depend on this choice.

For instance, if we consider the values reported in Ref.~\cite{miller09} for the 
black hole in Cygnus~X-1, this approach gives $E = 6.46$~keV as central value 
of the energy, while the lower and upper bounds are, respectively,
$E = 6.42$~keV (when $i = 27.9^\circ$ and $i_{\rm iron} = 29^\circ$) and
$E = 6.49$~keV (when $i = 26.3^\circ$ and $i_{\rm iron} = 31^\circ$).
So, $\Delta \alpha/\alpha = + 0.005 \pm 0.003$. The results in Ref.~\cite{duro11} 
give $E = 6.50$~keV as central value, and 
$E = 6.44$~keV (when $i = 27.9^\circ$ and $i_{\rm iron} = 30^\circ$) and
$E = 6.56$~keV (when $i = 26.3^\circ$ and $i_{\rm iron} = 34^\circ$) as,
respectively, lower and upper ones. The constraint on the fine structure constant is
$\Delta\alpha/\alpha = +0.008 \pm 0.005$.

If we want to include the uncertainty of the other parameters, we can proceed as
follows. For instance, including the effect of the uncertainty of the spin we can write
\be
\chi^2_{\rm red} (E_{\rm K\alpha}) &=&
\frac{1}{n-p} \sum_{i = 1}^{n} \frac{1}{\sigma^2_i} \Big[
N_i (E_{\rm K\alpha},\tilde{a}_* \pm \Delta\tilde{a}_*,\tilde{i}
\pm \Delta \tilde{i},\tilde{q},\tilde{r}_{\rm out}) + \nonumber\\ 
&&\hspace{1.8cm} - N_i (\tilde{E}_{\rm K\alpha},\tilde{a}_* \pm \Delta\tilde{a}_*,\tilde{i}_{\rm iron} 
\pm \Delta \tilde{i}_{\rm iron},\tilde{q},\tilde{r}_{\rm out}) 
\Big]^2 \, ,
\ee
and find the minimum of the reduced $\chi^2$. If we restrict the analysis to the very
small energy range including the peak and higher energies, we see that the effect of
the uncertainty on $a_*$ is roughly an order of magnitude smaller. The uncertainty
on $q$ and $\tilde{r}_{\rm out}$ also produces small effects. A proper error search 
over multiple parameters could be done better with multivariate gaussian draws 
from the parameter distribution.


\begin{acknowledgments} 
I thank Antonino Marciano, James Steiner, and Naoki Tsukamoto 
for useful comments and suggestions.
This work was supported by the NSFC grant No.~11305038, 
the Shanghai Municipal Education Commission grant for Innovative 
Programs No.~14ZZ001, the Thousand Young Talents Program, 
and Fudan University.
\end{acknowledgments}


\end{document}